\documentclass[aps,twocolumn]{revtex4}

\usepackage[dvips]{graphicx}

\renewcommand{\deg}{$^\mathrm{o}$}

\alph{footnote}

\begin{document}

\title{Growth and optical properties of GaN/AlN quantum wells}

\author{C.~Adelmann$^\mathrm{a)}$\footnotetext{$^\mathrm{a)}$ Present address: Department of Chemical Engineering and Materials Science, University of Minnesota, Minneapolis, MN 55455-0132}}
\author{E.~Sarigiannidou}
\author{D.~Jalabert}
\author{Y.~Hori$^\mathrm{b)}$\footnotetext{$^\mathrm{b)}$ On leave from NGK Insulators Ltd., Nagoya, Japan}}
\author{J.-L.~Rouvi\`ere}
\author{B.~Daudin$^\mathrm{c)}$\footnotetext{$^\mathrm{c)}$ Author to whom correspondence should be addressed; electronic mail: \texttt{bdaudin@cea.fr}}}
\affiliation{CEA/CNRS Research Group ``Nanophysique et Semiconducteurs'', DRFMC/SPMM, CEA/Grenoble, 17 Rue des Martyrs, 38054 Grenoble Cedex 9, France}

\author{S.~Fanget}
\author{C.~Bru-Chevallier}
\affiliation{Laboratoire de Physique de la Mati\`ere -- CNRS (UMR 5511), INSA de Lyon, 7 Avenue Jean Capelle, 69621 Villeurbanne Cedex, France}

\author{T.~Shibata}
\author{M.~Tanaka}
\affiliation{Corporate Technical Center, NGK Insulators Ltd., 2-56 Suda-cho, Mizuho, Nagoya 467-8530, Japan}

\begin{abstract}

We demonstrate the growth of GaN/AlN quantum well structures by plasma-assisted molecular-beam epitaxy by taking advantage of the surfactant effect of Ga. The GaN/AlN quantum wells show photoluminescence emission with photon energies in the range between 4.2 and 2.3\,eV for well widths between 0.7 and 2.6\,nm, respectively. An internal electric field strength of $9.2\pm 1.0$\,MV/cm is deduced from the dependence of the emission energy on the well width.

\end{abstract}

\maketitle

Group III nitride semiconductor compounds have recently demonstrated their capacity for light emission in the green to the near-ultraviolet (UV) spectral range \cite{Nakamura}. In the view of extending this range further into the UV region, GaN/(Al,Ga)N heterostructures have gained increasing interest. The optical properties of such heterostructures have been found to be strongly influenced by internal electric fields leading to a large red-shift of the emission wavelength by the quantum-confined Stark effect (QCSE) \cite{Im,Leroux,Widmann,Grandjean,Langer,Simon}. Such internal electric fields stem from both a piezoelectrical polarization due to biaxial strain in the nanostructure \cite{Im,Leroux,Widmann,Grandjean} or in the barriers \cite{Langer,Simon} as well as from the difference of the spontaneous polarization in the nanostructure and the barriers \cite{Bernardini97,Simon}.

The large majority of work in the GaN/(Al,Ga)N system has been performed for GaN/Al$_x$Ga$_{1-x}$N quantum wells (QWs), where the Al content $x$ is usually lower than about 0.3 \cite{Im,Leroux,Grandjean,Langer,Simon}. By contrast, little work has been devoted to GaN/AlN heterostructures \cite{Nam,Ohba,Keller}, with the exception of some reports on GaN/AlN quantum dots (QDs) \cite{Widmann,Damilano,Andreev,Simon2}. In the present letter, we report on the growth and optical properties GaN/AlN QWs and determine the value of the internal electric field for this material system, which can be used as a ``prototype'' system for the comparison of experimental and calculated magnitudes of internal electric field.

The samples were grown in a MECA2000 molecular-beam epitaxy growth chamber equipped with standard effusion cells for Al and Ga evaporation. The active nitrogen was provided by an Applied EPI Unibulb rf plasma cell. The plasma cell conditions were chosen to obtain GaN and AlN growth rates of 300\,nm/h under metal-rich (N-limited) conditions. The pseudo-substrates used were about 1.5\,$\mu$m thick (0001) (Al-polarity) AlN layers deposited by MOCVD on sapphire.

GaN (0001) (Ga-polarity) layers were deposited at a substrate temperature of 730\,\deg C. At this temperature, under near-stoichiometric or N-rich conditions, GaN QD formation following a Stranski-Krastanow growth mode has been reported \cite{Daudin}. However, it has also been shown that two-dimensional GaN growth can be obtained under strongly Ga-rich growth conditions due to the surfactant effect of excess Ga \cite{Mula}. For the growth of the GaN/AlN QWs studied in this letter, this effect was used to obtain two-dimensional flat GaN layers. They were subsequently rapidly overgrown by AlN under strongly Al-rich conditions before the evaporation of excess Ga on the surface (with an excess Ga film still present). The evaporation of this Ga film would lead to GaN islanding and thus to GaN QD formation \cite{Adelmann}. 

\begin{figure}[tb]
\includegraphics[width=7.5cm,clip]{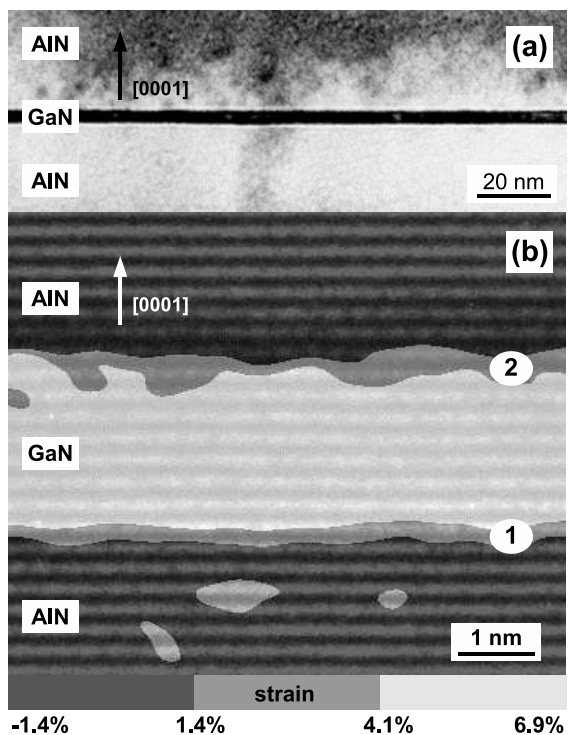}
\caption{\small \textbf{(a)} Bright-field cross-sectional TEM image [$\mathbf{g} = (0002)$] of a 2.6 nm wide GaN/AlN QW. \textbf{(b)} High-resolution TEM image superimposed to the result of a strain ($\epsilon_{zz}$) analysis. In the high-resolution TEM image, only (0002) planes are visible. The strain values have been divided into 3 intervals, as indicated. The width of interface 1 (GaN on AlN) is about 1\,ML, whereas interface 2 is $\sim 2$\,ML wide.}
\end{figure}

A cross-sectional transmission electron microscopy (TEM) image of a single 2.6\,nm thick GaN QW is depicted in Fig.~1(a). The GaN layer appears flat, regular, uniform, and exhibits no defects. One might expect that the excess Ga present on the surface at the beginning of AlN growth favor the formation of an Al$_x$Ga$_{1-x}$N alloy at the upper interface. However, a preliminary analysis of high resolution TEM images [Fig. 1~(b)] suggests that this process is rather limited, although the second interface (AlN grown on GaN) is slightly broader by about 1 monolayer (ML). This limitation is presumably due to preferential incorporation of Al with respect to Ga, leading to AlN growth even in the presence of Ga when Al is provided in excess of N. As a whole, interdiffusion appears restricted at this growth temperature, similar to the case of GaN/AlN QDs \cite{Arlery}. A more detailed TEM study will be published elsewhere \cite{Sari}.

The optical properties of GaN/AlN QWs were studied using a series of samples, each containing a single GaN QW grown capped by about 100\,nm of AlN. The QW width was varied between 0.7 and 2.6\,nm. The QW thickness was measured for all samples with a precision of $\pm 0.1$\,ML by \emph{ex situ} Rutherford backscattering spectrometry using 2\,MeV $\alpha$-particles from a van de Graaff accelerator. For some samples, the thickness was also cross-checked by TEM measurements and found in good agreement with the RBS results. Low-temperature (8\,K) photoluminescence (PL) was performed using a frequency-doubled cw Argon laser at $\lambda = 244$\,nm (5.08\,eV), \emph{i.e.{}} by pumping in the excited states of the QWs. The excitation density was a few W/cm$^{2}$. It is worth noting that the blueshift of the QW emission induced by increasing the excitation density by an order of magnitude was negligible with respect to the emission linewidth, which demonstrates that screening effects due to photo-induced carriers in the QWs can be precluded.

\begin{figure}[tb]
\includegraphics[width=8cm,clip]{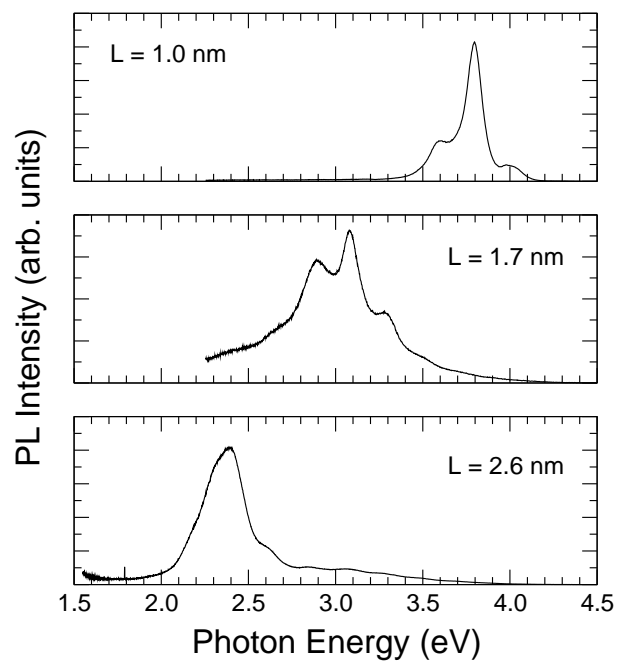}
\caption{\small Low-temperature PL spectra of three samples, each containing a single GaN QW in AlN barriers with widths $L$ as indicated.}
\end{figure}

PL spectra of GaN/AlN QWs for widths as indicated are depicted in Fig.~2. We observe that the emission energy decreases strongly with increasing QW width. At a QW thickness of around 1.3\,nm, the emission energy passes below the GaN bandgap energy, which can be assigned to the QCSE. For a QW thickness of 2.6\,nm, we observe PL emission at 2.35\,eV, \emph{i.e.{}} in the yellow spectral range. The full-width at half maximum (FWHM) of the PL peaks is around 300--350\,meV for all QWs and the oscillations of the PL intensity can be attributed to Fabry-Perot interferences between the sample surface and the AlN/sapphire interface.

The variation of the QW emission energy with the QW width is shown in Fig.~3. The solid lines represent the results of a calculation of the fundamental transition energies of a triangular QW. The GaN band gap was assumed to be $E_G (\mathrm{GaN}) = 3.645$\,eV under 2.4\% compressive biaxial strain \cite{Gil}; the AlN band gap was taken as $E_G (\mathrm{AlN}) = 6.28$\,eV, with a relative conduction band offset of $\Delta E_C/\Delta E_G = 0.75$ \cite{Martin}. The effective electron masses were assumed to be 0.22 and 0.40, the effective (heavy) hole masses 1 and 3.5 for GaN and AlN, respectively (in units of the free electron mass). Excitonic effects were neglected since an estimate showed that they excitonic binding energied are weak (a few 10\,meV) even for thin QWs due to the electric field-induced spatial separation of electrons and holes \cite{Andre}. To mimic the Stokes shift due to QW width fluctuations, $0.6\times \mathrm{FWHM}$ of the experimental PL linewidth was subtracted from the calculated transition energies (assuming that the adsorption linewidth is similar to the observed PL linewidth) \cite{Yang}. For simplicity, we use a constant offset of 200\,meV independent of the QW width. The internal electric field was used as a parameter. We find the experimental PL emission energies in good agreement with the calculations for an internal electric field between $F = 9$\,MV/cm and $F = 10$\,MV/cm. As the calculations show that the transition energy decreases linearly with increasing QW width, the electric field is more precisely determined by a linear fit to the data, yielding $9.2 \pm 1.0$\,MV/cm. With this value, note that the experimental FWHM of the PL peaks can be accounted for by monolayer fluctuations of the well width ($\pm 0.25$\,nm), in keeping with the HRTEM results. 

\begin{figure}[tb]
\includegraphics[width=8cm,clip]{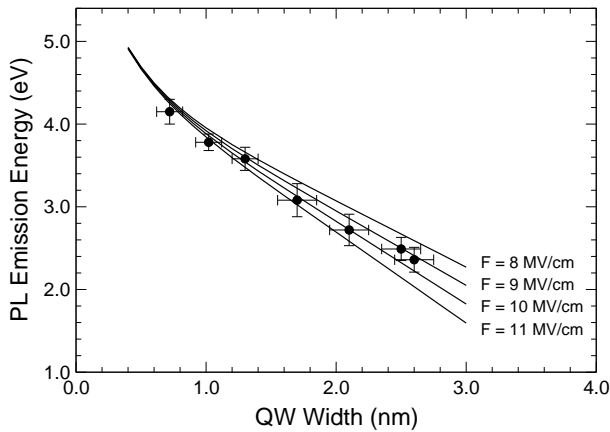}
\caption{\small Variation of the PL emission energy as a function of QW width. The solid lines represent calculations of the emission energy with internal electric field strengths $F$ as indicated.}
\end{figure}

The internal electric field in GaN/AlN QWs is slightly larger than that experimentally observed in GaN/AlN QDs (7\,MV/cm) \cite{Widmann}, which can be tentatively understood by a different strain state of QWs and QDs. The strain tensor of the latter is expected to contain a non-zero hydrostatic component, which has the tendency to lower the piezoelectric polarization.

The value of 9.2\,MV/cm is in excellent agreement with the theoretical prediction by Bernardini \emph{et al.} \cite{Bernardini97} of 9.5\,MV/cm for GaN/AlN. Note that our experimental result is a factor of 2 higher than the highest published theoretical value considering piezoelectric polarization alone (4.6\,MV/cm, Ref. \onlinecite{Martin}). This clearly demonstrates that piezoelectric polarization is not sufficient to explain the magnitude of the observed internal electric field and that spontaneous polarization is necessary to account for the observed magnitude of the internal electric fields in GaN/AlN QWs. This result further insinuates that the contributions of piezoelectric and spontaneous polarization are of the same order of magnitude.

We have also examined GaN/AlN QWs with width larger than 2.6\,nm, namely two samples with QW widths of 3.4 and 4.4\,nm, respectively. However, none of these samples showed PL emission within our experimentally-accessible spectral range down to about 0.7\,eV. An extrapolation of the fit in Fig.~3 shows that at least the 3.4\,nm wide QW should luminesce in this range. This suggests that plastic relaxation of the GaN layer might become significant around a thickness of $\sim 3$\,nm, \emph{i.e.{}} that, for larger GaN QWs, the mean distance between dislocations becomes of the order of the carrier diffusion length or the correlation length of the interface roughness (\emph{i.e.{}} the lateral confinement length scale), leading to a prevalence of non-radiative recombination.

In conclusion, we have demonstrated the growth of single GaN/AlN QWs by plasma-assisted molecular-beam epitaxy taking advantage of the Ga surfactant effect to induce two-dimensional GaN growth at a substrate temperature of 730\,\deg C. The GaN/AlN QWs show PL emission ranging from UV (4.2\,eV) to the yellow (2.3\,eV) spectral range. The variation of the QW emission energy with the QW width can be accounted for by an internal electric field of $9.2\pm 1.0$\,MV/cm. This unambiguously demonstrates that the optical properties of GaN/(Al,Ga)N heterostructures cannot be described in terms of a piezoelectric effect only and spontaneous polarization must also be considered.

The authors would like to thank R.~Andr\'e (Universit\'e Joseph Fourier, Grenoble, France) for providing the computer program to calculate the quantum well emission energies and N.~T.~Pelekanos (FORTH/IESL, Heraklion, Greece) for his careful reading of the manuscript.

\end{document}